\documentclass[a4paper,12pt,fleqn]{article}
\usepackage{latexsym}

\usepackage{amssymb}
\usepackage{amsmath}
\usepackage{amsfonts}

\begin{document}
\title{\bf Time-delay, energy-continuum, and systematics of particles}
\author{Sudhir R. Jain$^{(1)}$\footnote{srjain@apsara.barc.ernet.in}, 
Raja Ramanna$^{(2)}$, and K. Ramachandra$^{(2)}$\\
{\em $^{(1)}$Nuclear Physics Division, Van de Graaff Building,}\\ 
{\em Bhabha Atomic Research Centre, Trombay, Mumbai 400 085, India} \\
{\em $^{(2)}$ National Institute of Advanced Studies, I.I.Sc. Campus}\\
{\em Bangalore 560 012, India}}
\date{}
\maketitle

\begin{abstract}
Finding systematics in the mass-lifetime data for all the hadrons has 
been an outstanding problem. In this work, we show that the product of mass  
and lifetime for unstable 
particles is very well approximated by  $\hbar \frac{2^n}{n}$ where $n$ 
is an integer specific for a particle. In doing so, we have 
employed a relation between time-delay and resonances. The energy 
continuum has been treated in a way to take advantage of Cantor's 
mathematical work on continuum. Thus, 
even though the resonances are designated by a complex energy variable 
where ordering is not possible, in terms of stability, the index $n$ 
labels these resonances; larger the $n$, more stable a resonance is.  
\end{abstract}


\newpage
\section{Introduction}

One of the interesting problems of particle physics is to explain the 
relationship  between measured masses of fundamental particles and their 
lifetimes. There have been many attempts, of particular significance are 
those by Nambu \cite{nambu}, MacGregor \cite{macgregor}, and Ramanna 
\cite{ramanna}. Whereas in Nambu's work on empirical mass spectrum, the discreteness of particle masses was taken in terms of the mass of an electron, 
MacGregor's was in terms of the mass of a muon. In recent times, there have 
been a number of investigations connected to the  relation \cite{ramanna} 
which 
employs the measured masses and lifetimes of unstable nuclei and other 
particles :
\begin{equation}
\frac{MT}{\hbar} = \frac{2^n}{n}.
\end{equation} 
$M$ is the binding energy of the nuclei in $\alpha$-emitters, neutron mass 
in $\beta$-emitters, and the entire mass of decaying particles in case of 
fundamental particles. There is a compendium of the numbers `$n$' ensuing 
from the assumption of eq. (1). In this work, restricting to hadrons, 
we present a chain of arguments leading to (8), from where (1) follows as an excellent approximation.  
Considering that the hadron data is quite bizarre, finding a simple 
systematic rule through it is of great fundamental interest.  

There are several indirect reasons to convince us that the relation (1) 
holds. Of particular importance is the following : if we take the index of 
neutron, we can find the index $n$ for proton by employing the arguments 
in \cite{ramanna}. With this value of $n$, which turns out to be 225, the 
bound on lifetime of proton comes to be around $10^{33}$ years.   
This empirical relation (1)  has been used to obtain bounds on computational 
time and speed in quantum computers \cite{pjmr}. Also, the relation  
has inspired an interesting possibility in 
semiclassical chromodynamics  \cite{rrsrj}. 

There are many interesting instances in physics where concepts from the 
theory of numbers have been fruitfully employed. In treating the 
states in continuum 
in this work, we use some of the well-established results from Cantor's 
theory of transfinite numbers. This is quite unique and natural as the 
divisions in  energy can be made arbitrarily; however, it is interesting 
to make them with the hierarchy of infinities in mind. 

It is well-known  that a fundamental particle with a certain lifetime 
represents  a resonant state. Thus, to understand the origin of the 
empirical relation, (1), we have to start with a concept which is at the 
foundation of unstable or loosely-bound states - resonant states. 

In Section 2, we present a relation between time-delay and number of 
resonances below a certain energy. This relation was found in \cite{zasrj} 
where many examples have also been worked out. In Section 3, the continuum 
is treated with the help of Cantor's theory and by the help of an 
inequality from the theory of functions. In Section 4, we will find an 
index for each resonance and present our conclusions. 

\section{Time-delay and resonances}
   
We shall concentrate here on two-body scattering and two-body time-delay. 
The general definition of time-delay is given by the scalar product 
\cite{goldberger}: 
\begin{equation}
(f,Qf) = \lim_{R \to \infty} \int_{-\infty}^{+\infty} dt 
\left[(\psi (t),P(R)\psi (t)) - (\phi (t),P(R)\phi (t)) \right]
\end{equation}
where $f$ is an initial scattering state. $\psi (t)$ is the exact 
time-dependent wavefunction which asymptotically behaves as the freely 
evolving wavepacket $\phi (t)$ and $P(R)$ is given by 
\begin{eqnarray}
P(R) &=& 1, \mbox{~~inside the sphere of radius $R$} \nonumber \\
&=& 0, \mbox{~~outside the sphere of radius $R$}. 
\end{eqnarray}
From scattering theory, we know that 
\begin{eqnarray}
\phi (t) &=& e^{-iH_0t}f, \\
\psi (t) &=& e^{-iHt}\Omega ^{(+)}f,
\end{eqnarray}
where $\Omega ^{(+)}$ is the Moller operator. 

Before taking the limit on $R$, Eq. (2) represents the time difference 
that the exact wave $\psi (t)$ an the free wave $\phi (t)$ spend inside 
the sphere of radius $R$. The scalar product $(f, Qf)$ is the time-difference 
computed over all space. 

The problem of evaluating $(f, Qf)$ was eventually solved by Jauch and 
Marchand \cite{jauch}. We first state their result in the following. 
We can associate  a momentum space kernel $\langle p|Q|p'\rangle$ to 
$(f, Qf)$ :
\begin{equation}
(f, Qf) = \int f({\mathbf p})^*\langle {\mathbf p}|Q|{\mathbf p}'\rangle 
f({\mathbf p}') d{\mathbf p}d{\mathbf p}'.
\end{equation}     
$Q$ conserves energy and may be written as 
\begin{equation}
\langle {\mathbf p}|Q|{\mathbf p}'\rangle = \frac{\delta (E - E')}{\mu p}
\langle {\hat{\mathbf p}}|q(E)|{\hat{\mathbf p}}'\rangle ,
\end{equation}
where $q(E)$ is an operator on a two-dimensional Hilbert space. Energies are  
$E = p^2/2\mu$ and $E' = p'^2/2\mu$ where  $\mu$ is the reduced mass. 
Similarly, a reduced operator is introduced in \cite{jauch} for $S$-matrix 
elements through 
\begin{equation}
\langle {\mathbf p}|S|{\mathbf p}'\rangle = \frac{\delta (E - E')}{\mu p}
\langle {\hat{\mathbf p}}|s(E)|{\hat{\mathbf p}}'\rangle .
\end{equation}
The operator solution to the proposed problem is finally 
\begin{equation}
q(E) = -is^{\dagger}(E)\frac{d}{dE}s(E).
\end{equation}
In the above discussion, $\hbar$ is replaced by 1; however, in the 
last equation, it will appear multiplicatively if we re-insert it. 
This beautiful result was obtained in the rigorous work by 
Jauch, Sinha, and Misra \cite{jsm}. That this is a very general concept in 
scattering theory and is the same as sojourn times and global time delay 
has been reviewed in \cite{amrein}.  
 
There are two more developments which are important to mention to 
gain a correct perspective. From geometrical considerations, in 
classical and quantum kinematics, time delay was defined and its relation 
with $S$-matrix was found in \cite{heidi}. Subsequently, Narnhofer and 
Thirring \cite{nt} proved that the quasiclassical phase shift is a 
generator of classical canonical $S$ transformation. This result brought 
out a connection between resonances and looping trajectories, classical 
meaning of time delay and a classical meaning of Levinson's theorem. 
It is, in fact, from these works and using the techniques developed 
by Jauch and coworkers that it was proved that \cite{zasrj}: 
\begin{equation}
\int_{0}^{E^{\ast}} q(E) dE = n_R\hbar
\end{equation}
where $n_R$ is the number of resonances upto energy $E^{\ast}$. 

Tsang and Osborn \cite{to} came very close to this result but missed 
by just proving the Levinson's theorem in a different way. They showed 
for the $\ell $'th partial wave that 
\begin{equation}
\delta _{\ell}(0) = \pi n_B,
\end{equation}
where $n_B$ is the number of bound states. It is also written in terms 
of a difference between phase shifts at zero and infinite energy, a 
critique on this can be seen in \cite{zasrj}.  

It was noted in 
\cite{nt} that this wave-mechanical 
statement corresponds to a classical geometrical fact relating the volume in 
phase space of the bound orbits to the integral over the time delay. It is 
worth noting that this observation is the classical partner of Eq. (10). 
While Eq. (11) and hence the Levinson's theorem is concerned with 
bound states, Eq. (10) is concerned with resonances.    

This relation has also been used extensively in a recent work 
\cite{knkj} where all the nucleon and $\Delta$ resonances were extracted 
from experimental data.  

\section{Treating the continuum}

We have an integral relation in (10). This relation finds the number of 
resonances in  an interval. Given a continuum, finding a resonance is like 
searching a needle in the hay-stack. To locate it, we need to divide this 
interval in two parts initially, then sub-divide  both the divisions into 
two parts each, and repeat this process. In this way, depending on the width, 
a resonance will appear for some value of $n$, by then we would have 
made $2^n$ subdivisions of the original interval. For an index, $n$ on which 
a resonance exists, by Cantor's theory, $2^n$ is a continuum. It should be 
noted that we have brought in Cantor's concept as we are treating a continuum. 
Hitherto, these concepts have not been used in treating a continuous spectrum. 
However, as described above, this is also physically appealing. It is clear 
from the above discussion that a narrower resonance will be located on 
narrower bins and hence will lead to a large number of divisions, 
leading thereby to larger $n$. Let us now present it more rigorously.     

On making a discrete set on which the resonance exists, we need 
to incorporate this construction on (10) using some elementary 
property of the integral. To effect this, let us divide the interval 
$[0,E^{\ast}]$ into $2^n$ equal parts. Take $n$ of 
these consecutive parts and call these intervals 
as $J_1, J_2, \hdots , J_n$. (Omit the last 
interval as it may be fraction of other intervals.) Select any point, one in 
each of these intervals and call them $E_1, E_2, \hdots ,E_n$. 
Thus, we have divisions at $0$, $\frac{E^{\ast}}{2^n}$, 
$2$$\frac{E^{\ast}}{2^n}$, ..., $(n-1)$$\frac{E^{\ast}}{2^n}$. 
By a  property of the Riemann-Stieltjes integration, we have 
\begin{equation}
\bigg{|} n_R \hbar - \sum_{j=1}^{n} q (E_j)\Delta 
E_j \bigg{|} \leq ME^{\ast}~\mbox{max}_{j}~(\Delta E_j).
\end{equation}
$M$ is the maximum value of $\frac{dq}{dE}$. This will depend upon 
the distribution of lifetimes. 
By construction,  $\mbox{max}_{j}~(\Delta E_j)$ is 
$\frac{E^{\ast}}{2^n}$.

By re-arranging the inequality, we have
\begin{equation}
\frac{n}{2^n}E^{\ast} \geq \frac{n_R\hbar}{\sum_j q(E_j)} - 
\frac{1}{2^n} \frac{M {E^{\ast}}^2}{\sum_j q(E_j)}.
\end{equation}

\section{Stability index for a resonance}

Time delay gives a distribution of delay-times as defined through 
(2). We can obtain an expression for this distribution by posing 
the problem in the following interesting manner : a particle of 
given energy $E$ is scattered by a potential $V$ of range $R$. 
What is the mean time $T(E, a)$ that the particle spends inside a 
sphere of radius $r = a$, $a \geq R$ ? Solved in an ingenious way 
a long time ago \cite{baz64}, the solution for a single channel is 
\begin{equation}
T(E, a) = \frac{2}{v} \left[\frac{d\delta}{dk} + a - \frac{1}{2k}
\sin \{2(ka + \delta )\} \right].
\end{equation} 
If we make assumptions of the state being a quasistationary state 
\cite{baz}  
and thus referring to a resonance at an energy $E_0$ with width $\Gamma$ 
so that the phase shift is 
\begin{equation} 
\delta = \varphi - \tan ^{-1} \frac{\Gamma /2}{E - E_0},
\end{equation}
upto the leading term, 
\begin{equation}
T(E) = \frac{\hbar \Gamma}{(E - E_0)^2 + (\Gamma /2)^2}.
\end{equation}
In fact, $T(E)$ is the same as $q(E)$. Note that (14) corresponds 
to an average time $\overline{T}$ that the particle spends inside 
a sphere of radius $r=a$. Thus, we can even consider a distribution 
of $T$'s. Remarkably, it has been shown \cite{baz} that this 
probability distribution function is 
$\sim \delta (T - \overline{T})$. This does not clash with the 
uncertainty relation $\Delta E\Delta t \geq \hbar$ as we know that 
while considering the motion of wavepackets, $\Delta t$ is the 
uncertainty at the collision instant; it has no relation with the 
duration of the collision or interaction.     
   
For an individual resonance, (13) will modify simply as 
\begin{equation}
\frac{n_0}{2^{n_0}}E^{\ast} \geq \frac{\hbar}{q_0} - 
\frac{1}{2^{n_0}} \frac{M E^{\ast}\Gamma}{q_0},
\end{equation}
where for simplicity, we have assumed that the resonances are 
quasistationary states corresponding to the second sheet of the Riemann 
surface of $s(E)$. The maximum of 
$\frac{dq}{dE}$ occurs at $E_0 - \frac{\Gamma}{2\sqrt{3}}$. The 
maximum $M = \frac{3\sqrt{3}\hbar}{\Gamma ^2}$.   
Eq. (17) can now be written as 
\begin{equation}
\frac{E^{\ast}\tau _0}{\hbar} \geq \frac{2^{n_0}}{n_0}
\left[1 - \frac{3\sqrt{3}}{2^{n_0}}\frac{E^{\ast}}{\Gamma} \right]. 
\end{equation}
We can now compare the index $n_0$ coming from (1) \cite{rs} with 
that coming 
from the positivity of the quantity in the square bracket in (18). 
Thus, we must have $n_0$ satisfying 
\begin{equation}
n_0 \geq \frac{\log (3\sqrt{3}E^\ast /\Gamma )}{\log 2}.
\end{equation}
Given the energy $E^\ast$ and width, we have now an inequality 
binding $n_0$ for a resonance. In the next Section, we will compare 
the values coming from (19) with those found assuming (1). 

\section{Systematics in terms of the stability index}

In Table 1, we present the comparison of the values for $n_0$ following (19) 
with the ones found empirically in \cite{ramanna}.  It is amazing that 
(19) almost reproduces the compendium 
of `$n_0$'s ensuing from the assumption of (1). This remarkable agreement 
implies that with the index $n$ taken as the first integer greater than 
(19), $2^n/n$ reproduces the dimensionless  ratio $MT/\hbar$. This proves 
(1) and hence provides a global systematization of masses and lifetimes 
of all the hadrons in such a simple way. For the $K^+$ meson, (19) does not 
predict the correct index, and this is not clear to us. However, this is 
the only exception known to us; for the rest, at least the order of magnitude 
is correct. This becomes more important as we have presented a proof for 
the relation conjectured earlier.        
\vskip 0.25 cm
\begin{table}[t]
\caption{\it A comparison between the index $n$ obtained by assuming Eq. (1) and one deduced from our analysis (Eq. (19) for an arbitrarily chosen collection of hadrons. The remarkable agreement is seen here over 27 orders of magnitude in the widths of resonances serves as a test of Eqs (1) and (18,19). For almost all the hadrons known to us, the agreement is as good.}

\begin{tabular}{|l|c|c|c|c|}

\hline
hadron & Mass (MeV) & Width (MeV) & $n$ (Eq. (1)) & $n_0$\\
\hline
$n$ & 939 & 7.43 $\times$ $10^{-25}$ & 97 & 93 \\
\hline
$\Lambda$ & 1120 & 2.5 $\times$ $10^{-12}$ & 54 & 52 \\
\hline
$B^0$ & 5280 & 4.39 $\times $ $10^{-10}$ & 49 & 46\\
\hline
$J/\psi (1S)$ & 3100 & 8.8 $\times$ $10^{-2}$ & 19 & 18 \\
\hline
$\chi ^{c1} 1P$ & 3510 & 8.8 $\times$ $10^{-1}$ & 16 & 15 \\
\hline
$D_{s1}(2536)^{\pm}$ & 2536 & 4.5 & 13 & 12\\
\hline
$\psi (4415)$ & 4415 & 43 & 10 & 10 \\
\hline
$\Xi (1820) ~D_{13}$ & 1820 & 24 & 9 & 9 \\
\hline
$\Lambda (1690) ~D_{03}$ & 1690 & 60 & 8 & 8 \\
\hline
$\Sigma (1750) ~S_{11}$ & 1750 & 110 & 7 & 7 \\
\hline
$N(1520)~D_{13}$ & 1520 & 123 & 6 & 7 \\
\hline
$\Delta (1232) ~P_{33}$ & 1232 & 120 & 6 & 6\\
\hline
\end{tabular}
\end{table}
The remarkable agreement of the values of stability indices found in 
Table 1 suggests that there must be a proof of (1), or, that there 
exists a relation closely related to (1). To see this, almost trivially 
now, we employ  the relation, $\Gamma = \hbar/\tau _0$ and  get 
\begin{equation}
\frac{E^\ast \tau _0}{\hbar} \geq \frac{2^{n_0}}{n_0 + 3\sqrt{3}} 
= \frac{2^{n_0}}{n_0}
\left[1 - \frac{3\sqrt{3}}{n_0} + \left(\frac{3\sqrt{3}}{n_0}\right)^2 - 
\cdots \right].
\end{equation}
From this, it follows that the empirical relation (1) is a good approximation 
to the above result  for the cases when $n_0 > 3\sqrt{3}$. Most of the hadrons 
belong to the situation where $n_0 > 5$. Eq. (19) is an inequality, thus there 
could be  many $n_0$'s. We will consistently take the first integer larger 
than $n_0$ following (19) for each hadron. This entails the compendium 
of the index, $n_0$.  

\section{Remarks}

We wish to make some important points here which clarify some subtle 
issues underlying the discussion presented above. These points also 
show some limitations of our approach and suggest future directions 
which could be of great interest. 

\begin{enumerate}

\item According to the ``bootstrap'' hypothesis, all strongly interacting 
particles are bound states or resonances \cite{rfd}. We have employed 
the $S$-matrix method and in this approach, closely related methods 
apply to both nonrelativistic and relativistic problems. One can use 
the understanding of bound states, resonances, and perturbations on the 
interaction that one has in nonrelativistic quantum mechanics as a guide 
in relativistic problems which, according to the ``bootstrap'' hypothesis, 
possess these features. 

\item In the $S$-matrix formulation of statistical mechanics, the change 
in particle number is taken care of. The $S$-matrix formulation helps in 
extending the nonrelativistic statistical mechanics to relativistic 
statistical mechanics. We have taken the intermediate state of a particle 
(e.g., proton) and another particle (say, a field quantum like $\pi ^+$ meson) 
as a resonance because it is neither a bound state nor a scattering state. 
When applied to nucleon and $\Delta$ resonances, we have been successful in 
extracting correct partial widths for corresponding channels \cite{knkj}.

\item We have only considered two-body time-delay. There could be multiparticle 
time-delay involving more than two particles. We do not know the effect of 
those details except that they effects will only be at a less significant 
order. However, we emphasize that there are examples like $\omega$ meson 
which decays in $\pi ^+, \pi ^-, \pi ^0$ mesons with a probability 0.88; 
these particles will be explainable only by employing three-particle 
time-delay \cite{bo}.      

\item For most of the particles, there are many channels of transformation. 
In the multi-channel case, the time-delay used here and in other related 
works is only correct in a quasi-classical approximation. This result is 
not well-known and can be seen in a footnote in \cite{baz}. 

\item  We have assumed a Lorentzian distribution for time-delay, whereas it 
is recently shown that some of the hadron  resonances   may not have 
a Lorentzian shape \cite{jat}. There are also overlapping resonances. We 
have made simplifying assumptions. 

\item It is possible to choose the division of the interval in (4) so that 
instead of $\frac{2^n}{n}$, some other form like $\frac{2^{f(n)}}{f(n)}$ 
ensues. However, that would not make any conceptual difference. Also, 
here $2^n$ is chosen as an expression of the continuum in comparison with 
$n$, an expression naturally coming from Cantor's well-known theory.  
This is the most natural choice. Thus, we have been able to label each 
resonance by an integer, $n$, which is remarkable. 
Also, note that larger the $n$, more stable the particle is. We have 
called this index as a ``stability index''.

\item We believe that the relation found here would have a microscopic origin. 
It is already clear that it is consistent with quantum scattering theory 
(relativistic and non-relativistic) as we started with the relation (3). 
\end{enumerate}

\vskip 1.0 truecm

\noindent
{\bf Acknowledgements}

SRJ thanks Zafar Ahmed and A. K. Jain 
for important discussions. SRJ also thanks the National Institute of 
Advanced Studies for warm hospitality extended towards him during his 
visits.

\end{document}